\documentstyle[12pt,aasms4] {article}

\begin{document}

\title{\bf NO MASSIVE BLACK HOLE IN CYG X-3}

\author{\it Abhas Mitra\altaffilmark{1}}

\altaffiltext{1}{Theoretical Physics Division, Bhabha Atomic Research Center,
Mumbai- 400 085, India\\ E-mail: nrl@magnum.barc.ernet.in \\also
krsrini@magnum.barc.ernet.in}

\begin{abstract}
There has been a recent suggestion that Cyg X-3 contains a black hole (BH)
of mass $M_1 \sim 17 M_{\odot}$. This interpretation is closely linked to
a previous claim that Cyg X-3 contains a Wolf-Rayet star of mass $M_2
\sim 10 M_{\odot}$. The latter interpretation would imply that the X-ray
source in this close binary ($P = 4.8$ hr) is enshrouded by a relatively
cool superstrong wind with ${\dot M} \ge 10^{-5}~ M_{\odot}$~  yr$^{-1}$.  It
can be shown that such a wind would be completely opaque to the low energy
X-rays observed from the source and hence the Wolf-Rayet hypothesis can
not be a correct one. By pursuing the same argument it also follows that
the compact object must be of modest mass ruling out the existence of a
massive BH in Cyg X-3. Note, however, that  at present we can not strictly
rule out the probable  existence of stellar mass BH in Cyg X-3 or any
other X-ray binary which are believed to contain a neutron star as the
compact object.
Yet, a more probable scenario for Cyg X-3
would be one where the compact object is a canonical neutron star (NS) and
the companion is an extremely low mass dwarf, $M_2 \sim 10^{-2}~
M_{\odot}$ much like $PSR ~1957+20$.

\end{abstract}


\section{INTRODUCTION}
 In a recent paper, Schmutz, Geballe \& Schild (1996, henceforth SGS) have
claimed to have presented evidence for the existence of a $\sim
17~M_\odot$ BH in Cyg X-3.  This conclusion is based primarily on the
previous interpretation of van Kerkwijk et al. (1992, VK) that the
companion of Cyg X-3 is a moderately massive Wolf-Rayet star of mass, $M_2
\sim~10~M_\odot$.  SGS have claimed  to have measured the value of the
mass function of the binary as $f(m) = 2.3~M_\odot $ by using the
numerical relationship
\begin{equation}
f(m) = 1.035\times 10^{-7} K^3 P(1-e^2)^{3/2} = {M_1^3 \sin^3i \over
{(M_1 + M_2)^2}}
\end{equation}
\par where $P$ is the binary period in days, $e$ is the eccentricity of
the binary, $i$ is the angle of inclination of the orbit, and $K$ is the
measured orbital Doppler velocity amplitude (in Km s$^{-1}$).
SGS obtained $M_1 = 17 M_{\odot}$
by taking a value of $e=0$, $P=0.2$, $i=50^o$, $K = 480 \pm 20$~ Km
s$^{-1}$, and $M_2$ (Wolf-Rayet)$= 13~ M_{\odot}$.  However, in a
recent paper (Mitra 1996, henceforth M96), it has been discussed in
considerable detail why this Wolf-Rayet interpretation is unlikely to be
true for Cyg X-3. Furthermore, a careful scrutiny of SGS would reveal that
the velocity amplitude measured by
SGS may not be related to orbital motion at all.
 The former point alone would show that even if we take the quoted
value of $K$, the compact object may not have a mass higher than $2.7
M_{\odot}$.

\par M96 attempted to show that had Cyg X-3 really contained a Wolf-Rayet
star with a wind much stronger than $10^{-7} M_\odot$ yr$^{-1}$, it would be
opaque to low energy X-rays. And massive Wolf-Rayet stars have wind
stronger than $\dot M \ge 10^{-5} M_\odot$ yr$^{-1}$. Following the
analytical work on photoionization opacities of a cosmic plasma by
different authors, it was
shown in M96 that for $2-8$ keV X-rays
passing through a wind, the photoelectric absorption cross-section
approximately varies as $\sigma~\sim~5\times 10^{-22} (E/1~{\rm keV})^{-3}$
cm$^2$ per H atom.  Simultaneously the thick wind offers a very large column
density $(l)$ to the central X-ray source which results in a very large
absorption optical thickness $ \tau~>>~1 $ for the low energy X-rays.

\par However, there is a scope for confusion in this regard because Teresawa \&
Nakamura (1993) have claimed to have applied the CLOUDY photoionization code to
the model of Cyg X-3 to find that the Wolf-Rayet wind need not be opaque
to the X-rays. In M96, we discussed in detail why this claim can not be
true. One of the reasons being that one can not conceive of a wind whose
metallicity is lower than the basic cosmic composition; in fact, the wind
of all evolved stars are likely to have metallicities much higher than the
basic cosmic values. Correspondingly, the actual opacities of the W-R wind
can be much higher than what has been estimated in M96.  And it is important to note that
the fact that
a wind as strong as that of a  Wolf-Rayet star,  even if it were made of
cosmic composition, would be absolutely opaque to soft X-rays for a close
X-ray binary with $ P=4.8$ hr
 {\em has been verified independently by
A.C. Fabian  by
making use of the same photoionization code CLOUDY} (M96).
Since the resultant optical depths could be as large as $\sim 10^3$,
probable uncertainties in such calculations would not change the basic
conclusion that a Wolf-Rayet wind would be absolutely opaque to soft X-rays
atleast for a close binary like Cyg X-3.

\par Nonetheless, in M96, it was assumed that the compact object has a mass
 $ \sim 1.4M_{\odot} $ and let us try to adjudge the probable changes in
the former interpretation in case the value of $(M_1 + M_2)$ were $\sim 30
M_{\odot}$. Since the semimajor axis of the binary changes only modestly
in this process, $ a
\sim~(M_1 + M_2)^{1/3} $, the value of $l$ is lowered approximately
by a factor of three for a given value of $\dot M$ and mean wind speed
$v$. On the other hand, the average value of the ionization parameter,
$\xi$, (Tarter, Tucker \& Salpeter 1969) remains approximately constant:
\begin{equation}
\xi = {L_x \over {n r_x^2}} = {{4\pi v m r^2 L_x}
\over {{\dot M} r_x^2}}
\end{equation}
\par where $L_x$ is the X-ray luminosity, $n$ is the atomic number
density, $r$ is the distance measured from the center of the companion and
$r_x$ is the distance measured from the center of the compact object. For
a mean value of ${r \over r_x}\sim1$, we can see that  the value of $\xi$
does not change significantly. In M96, the value of $v$ was taken to be
1000 Km s$^{-1}$. Even if the value of $v$ is higher, say, 2500 Km s$^{-1}$,
 for the extremely high value of $\dot M > 10^{-5} M_{\odot}~
$yr$^{-1}$ required by SGS, we would still have a low value of $\xi \sim
10^2$ as was originally found in M96.

\par Consequently, the value of the electron temperature of the wind
continues to be very low $\sim 10^4 K$. Since, for a cosmic plasma, the
previously referred value of  the photoelectric absorption cross-section
$\sigma~\sim~5\times 10^{-22} (E/1~{\rm keV})^{-3}$ cm$^2$~ per H-atom  holds
good over a wide range of temperature $
\sim~10^4-10^6$~ K, all that happens now is that the final optical depths
would be reduced by a factor of $3 \times 2.5 \sim 7.5$. This means that
if the claim of SGS were true, still we would have values of $\tau$
crudely ranging between $40 - 400$ depending on the value of $i$. Thus,
the actual value of $\dot M$ must be much lower than what is implied by
SGS. And simply this fact rather than any presumed value of $M_1$ would
try to constrain the value of $M_2$.

\par  Therefore, we believe that the conclusion of M96 that $
q~\equiv~{M_2/M_1} ~<<~1 $ for Cyg X-3 remains valid for a wide range of
values of $M_1$ and $M_2$.

\section{ REANALYSIS}
  There are several implicit assumptions behind the philosophy of
determining the mass function of a the companion(s) of a binary.  One
assumption is that the region from where the probe line or signal is
emitted behaves like a point source and tracks the orbital motion.
 In otherwords the radial extent of
the region $\Delta r \ll a$, the semimajor axis of the binary.  This
 condition is
approximately satisfied for optical lines emitted from the surface
(photosphere) of even a massive star orbiting a wide binary ($P \gg 1$ d),
or for emission from a compact object like a neutron star in case of
binary (radio) pulsars or X-ray binaries even when the value of $P$ is
only a few hr. And if the IR lines observed by SGS really emanate from the
surface of the companion in Cyg X-3,this condition might be valid.
Even then, for $q \ll  1$, we obtain from equn. (1) $f(m) \sim M_1 \sin^3 i$.
Note here that, so
far, the best observational limit on the inclination angle is that due to
White \& Holt (1982) and which was obtained from the X-ray studies of Cyg X-3:
$i \le~70^o$. Therefore, a value of $f(m) =2.3~ M_{\odot}$ {\em could actually
imply  $M_1 \approx 2.7~M_\odot$ contrary to the estimate of SGS}, $M_1
\approx 17~M_\odot$!  The compact object of Cyg X-3 is likely to be
spinning rapidly and therefore a value of $M_1 \approx 2.7~M_\odot$ is
well below upper mass limit of a similar neutron star ($\sim 3.8~M_\odot$)
(Friedman \& Ipser 1987). This range of a value of $M_1$ is in broad
agreement with the general idea that the compact object in Cyg X-3 is
radiating at a near Eddington rate (Mitra 1992a).

\par  The above conclusion that $M_1 \sim 2.7 M_{\sun}$ has been reached by considering the
interpretation of SGS that the measured velocity dispersion could be
really ascribed to the orbital motion. However, following some
clarification due to  R. Gies (personal communication), now we point out that this
very interpretation is unlikely to be true because of the following
reasons. In their paper, SGS show a light curve in their Fig. 1 and a
radial velocity curve in their Fig. 5. We see that the IR flux minimum
(which must occur at an orbital conjunction phase) occurs at the same epoch as the
radial velocity extremum, which, if it were, the result of orbital motion,
must occur at a quadrature phase. Thus {\em it is unlikely that SGS  have
  measured the
orbital motion}.
Instead, it appears much more probable that that the features SGS have
measured are related to the structures in the wind outflow. In fact, their
light curve and line profiles resemble the models for wind outflow given
by van Kerkwijk et al. (1996), and thus it seems extremely likely that the
SGS velocity curve reflects changes in the orientation of structure in the
wind and not orbital motion. Note also that in any case, the observed
lines are supposed to be produced far off in the wind in a region
which is relatively cool ($\Delta r \gg a$?). If this is true, the entire work of SGS is rendered
irrelevant for the use of mass-function formula (1).

\par Let us remind here that if the companion is a He-rich dwarf,  then
strong X-ray irradiation due to the compact object may ablate the
companion and generate an evaporative wind either from the companion or
from the accretion disk or from both. And in such a case, we would not be
hard put to explain how we can have a W-R  star with a radius apparently larger than
the narrow Roche lobe of Cyg X-3 as the mass donating star. Note also that this
latter difficulty is usually explained away by assuming that the strong
emission lines are emitted from a region much larger than the orbit size (
$\Delta r \gg a$).
If it were really so, again, we are led to the conclusion that the
observations of SGS do not refer to the orbital motion at all.

\section {CONCLUSION}
 We find that even if we take the measured value of SGS at its face
($K=480 \pm 20$ km s$^{-1}$), the value of the mass of the compact object
in Cyg X-3 could be as low as $2.7 M_\odot$ and which invalidates the
claim for finding evidence for a massive $\sim 17~M_\odot$ BH in Cyg X-3.
However, it should be borne in mind, that in a strict sense,  at
present, one cannot absolutely rule out the existence of a {\em stellar
mass} BH in Cyg X-3 or for that matter in most of the low mass X-ray
binaries which are normally believed to contain a neutron star as the
compact object.  Yet more importantly,  we find that the measurement
of SGS may not at all be ascribed to orbital motions rendering it
irrelevant for determination of the mass function(s) of the binary.

\par It may be worthwhile to recall the backdrop against which this idea of
having a massive BH in Cyg X-3 arose; obviously it was the suggestion put
forward by van Kerkwijk et al. (1992) that the detected He-lines in the
infrared spectrum of Cyg X-3 could be due to a massime Wolf-Rayet star.
 Note, however, that, in the same paper, it
was mentioned that

``Such a wind might be due X-ray irradiation of either the accretion disk
or a low mass companion. {\em Alternatively}, it could be intrinsic to the
companion, that is the companion could be a WR star. For lack of
prediction, the irradiation models are extremely difficult to test. Below,
we therefore cofine ourselves to the WR model.''

Unfortunately, it did not take long to forgo the  reasonableness expressed 
in the foregoing statement and many researchers took the more fashionable
Wolf-Rayet hpothesis to be granted by completely ignoring the wind opacity
problem.
 Once, we accept that the the compact object is of modest mass and use the
fact that $q\ll 1$, the analysis of M96 would strongly suggest
the companion in this binary is a very low mass object $M_2
\sim 0.01 M_{\odot}$ and, presumably, a He-rich white dwarf ablated in
response to the bombardment of various radiations emanating from
the compact object (Mitra 1992b).
This given estimate would be in agreement with the idea
that Cyg X-3 could be an immediate predecessor of a system like PRS 1957+20.
Then the evolutionary status of Cyg X-3 suggested by the present work and
M96 would be quite similar to the one discussed by Tavani, Ruderman, \&
Shaham (1989).

However, we must bear in mind that the  evolutionary model of Cyg
X-3 which has been suggested here or any other model suggested by other authors 
must be considered tentative at the present stage. Only when such models are
developed further and it will be found that they are  capable of reproducing the
observed He and other line strengths, 
  such models may be considered to be
realistic ones.

\noindent Acknowledgement: The author wishes to thank R. Gies for
illuminating discussions on several aspects of the present work.

\end{document}